\documentstyle[preprint,aps,tighten,epsf]{revtex}

\begin{document}
\draft 

\preprint{\it 17th Nordic Semiconductor Meeting, Trondheim (to appear
in Physica Scripta)}

\title{Parametric correlation of Coulomb blockade conductance peaks in
chaotic quantum dots}

\author{Henrik Bruus}

\address{CNRS-CRTBT, 25 Avenue des Martyrs, BP166, F-38042
	 Grenoble C\'edex 9, France}

\author{Caio H. Lewenkopf}

\address{Instituto de F\'{\i}sica, UERJ, R. S\~ao Francisco Xavier, 524,  
	 20559-900 Rio de Janeiro, Brazil}

\author{Eduardo R. Mucciolo}

\address{NORDITA, Blegdamsvej 17, DK-2100 Copenhagen {\O}, Denmark}

\date{May 15, 1996}

\maketitle

\begin{abstract}
We investigate the autocorrelator of conductance peak heights for
quantum dots in the Coulomb blockade regime. Analytical and numerical
results based on Random Matrix Theory are presented and compared to
exact numerical calculations based on a simple dynamical model. We
consider the case of preserved time-reversal symmetry, which is
realized experimentally by varying the shape of the quantum dot in the
absence of magnetic fields. Upon a proper rescaling, the correlator
becomes independent of the details of the system and its form is
solely determined by symmetry properties and the number of channels in
the leads. The magnitude of the scaling parameter is estimated by a
semiclassical approach.
\end{abstract}

\draft

\narrowtext

\section{Introduction}
\label{sec:intro}

Since the first experiments dealing with coherence phenomena in
microstructures in the 80's and the theoretical activity that
followed, many experiments have probed several different universal
features of mesoscopic devices. One important example of such systems
are semiconductor quantum dots \cite{Kastner92}, the subject of our
study. With the fast progress in nanotechnology, the experiments
gained a lot in sophistication and were able to explore more subtle
aspects of mesoscopic physics. One of the most interesting subjects in
the field is related to the influence of chaos in the ballistic
electronic transport
\cite{ChaosReview,Jalabert92,Prigodin93,Bruus94,Mucciolo95a,Alhassid95,Alhassid96,Bruus96}.

Presently, we know that the statistical measures of conductance
(distributions and correlators) in open systems are well described by
universal curves obtained from Random Matrix Theory (RMT). However,
the understanding of system-dependent properties (average conductance,
correlation scales of external parameters, etc.) demands a
complementary approach, which extracts relevant elements of the
underlying classical dynamics or average quantum properties of the
system at hand. In particular, the decay width of the conductance
correlators as a function of either an external magnetic field or of
variations in the shape of the quantum dot can be understood in
semiclassical terms.

In this work we extend our recent analysis \cite{Bruus96} of the
correlator of Coulomb blockade conductance peak heights for quantum
dots in the chaotic regime. Here we concentrate on the case of
preserved time-reversal symmetry and generalize our previous
semiclassical analysis. We use the same notation as in our previous
paper.

The conductance $G$ of mesoscopic devices, such as quantum dots, can
be evaluated using the Landauer-B\"uttiker formula
\cite{Landauer57,Buettiker}. For quantum dots the latter becomes
particularly simple and can be expressed in terms of the partial decay
widths $\Gamma_{c\nu}$ and partial decay amplitudes $\gamma_{c\nu}$
\cite{Jalabert92}:
\begin{equation}
\Gamma_{c\nu} = |\gamma_{c\nu}|^2,
\qquad {\rm with} \qquad
\gamma_{c\nu} = \sqrt{\frac{\hbar^2}{2m}} \int
\!ds\,\chi_c^*(\bbox{r})\psi_\nu(\bbox{r})\;,
\label{eq:gamma}
\end{equation}
where $\psi_\nu$ is the eigenfunction of the resonance $\nu$ {\it
inside} the dot with the appropriate boundary condition and $\chi_c$
is the wave function at energy $\varepsilon_\nu$ for the channel $c$
belonging to one of the leads. The integration is performed over the
contact region between the lead and the quantum
dot. Equation~(\ref{eq:gamma}) does not include barrier penetration
factors \cite{Bruus96}, which should not be important for the case of
one open channel in each lead -- the case we analyze here and the one
which is presently more relevant for comparison with experiments. In
this framework, the influence of finite temperatures caused by the
rounding of the Fermi distribution can be done straightforwardly. In
the single-level limit $\Gamma_{c\nu}\ll kT<\Delta$ ($\Delta$ is the
single-particle mean level spacing), the conductance peak height
$G_\nu$ corresponding to an {\it on resonance} measurement is given 
by \cite{Beenakker91}
\begin{equation}
\label{eq:peakcond}
    G_\nu = \frac{2e^2}{h} \left( \frac{\pi}{2kT} \right) g_\nu \;,
	  \qquad \mbox{with} \qquad g_\nu = \frac{\Gamma_{L\nu}
	  \,\Gamma_{R\nu}} {\Gamma_{L\nu} + \Gamma_{R\nu}} \; ,
\end{equation}
where $\Gamma_{L(R)\nu}$ is the width of the resonance $\nu$
corresponding to a decay into open channels in the $L(R)$ lead. In
other words, $\Gamma_{L(R)\nu} = \sum_{c\in L(R)} \Gamma_{c\nu}$. If
the quantum dot has any asymmetry (or if there is a small amount of
disorder) and the leads are placed far from each other on the scale of
the electron wavelength, $\Gamma_{R\nu}$ and $\Gamma_{L\nu}$ will not
be identical and should, in fact, vary independently.

\section{Parametric correlators and Random Matrix Theory}
\label{sec:paracorRMT}

We are interested in universal properties of a system Hamiltonian $H$
which depends on an external parameter $X$ and whose underlying
dynamics is chaotic. Without loss of generality, we shall illustrate
our study by considering $X$ as a measure of the shape deformation of
the quantum dot boundaries (in Ref.~\cite{Bruus96} the parameter $X$
was identified with the magnetic flux threading the quantum dot).
Furthermore, we assume that in the absence of a magnetic field, $H$
can be modeled as a member of the Gaussian orthogonal ensemble (GOE)
of random matrices \cite{Mehta91}. Thus, the statistical properties of
the system should be obtained by an ensemble averaging. We focus our
study on two autocorrelation functions: the level velocity correlator
$C_v(X)$ and the conductance peak correlator $C_g(X)$.

\begin{figure}
\setlength{\unitlength}{1mm}
\begin{picture}(140,140)(0,40)
\put(0,0){\epsfxsize=14cm\epsfbox{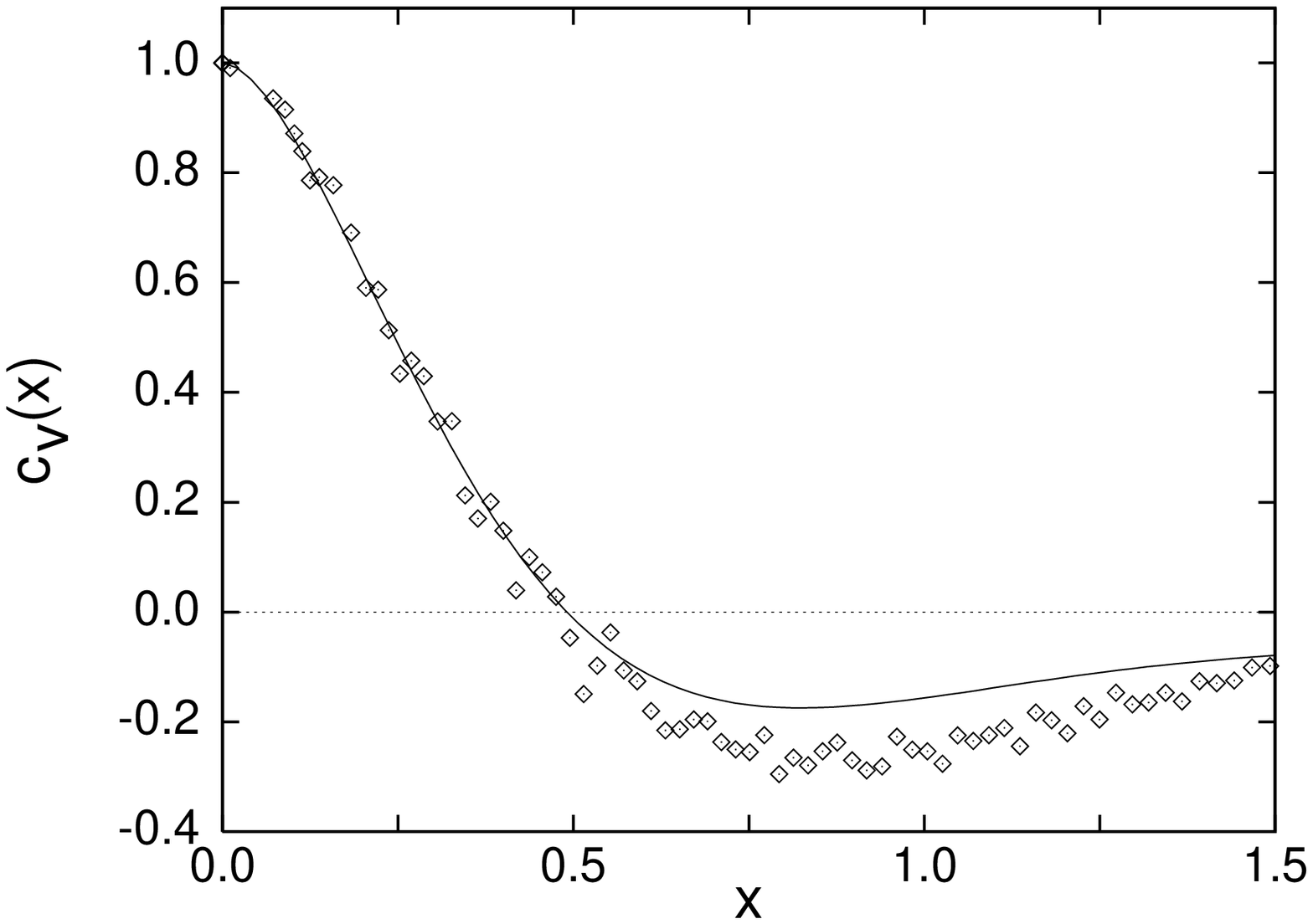}}
\end{picture}
\caption{The rescaled level velocity correlator $c_v(x)$ as a function
of shape deformation. The full line is the RMT result for the GOE
matrix diagonalizations while the points correspond to exact numerical
calculations for the conformal billiard.}
\label{fig:CvX}
\end{figure}

\begin{figure}
\setlength{\unitlength}{1mm}
\begin{picture}(140,140)(0,40)
\put(0,0){\epsfxsize=14cm\epsfbox{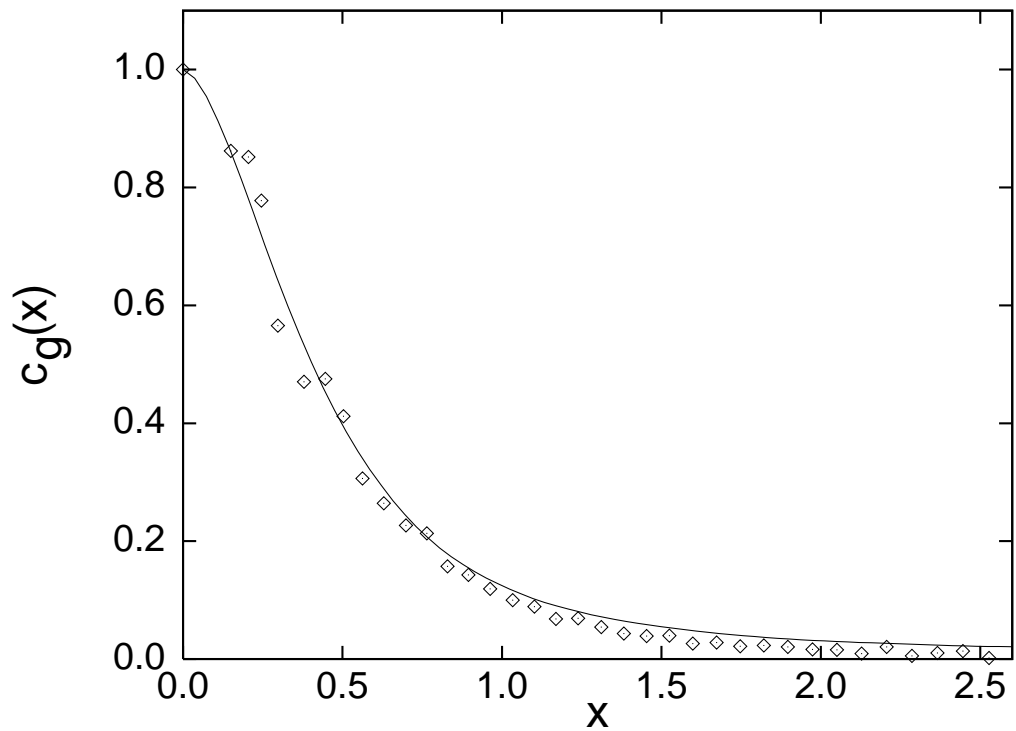}}
\end{picture}
\caption{The conductance peak height correlator $c_g(x)$. The symbols
follow the same conventions of Fig.~1.}
\label{fig:CgX}
\end{figure}

As originally found by Szafer, Altshuler, and Simons
\cite{Szafer93,Simons93a} the level velocity correlator
\begin{equation}
\label{CvXdef}
C_v(X) \equiv \frac{1}{\Delta^2} \left[ \left\langle
\frac{d\varepsilon_\nu(\bar{X}-X/2)}{d\bar{X}}\ \frac{d\varepsilon_\nu
(\bar{X}+X/2)}{d\bar{X}}\right\rangle - \left\langle
\frac{d\varepsilon_\nu(\bar{X})}{d\bar{X}} \right\rangle^2 \right] ,
\end{equation}
is a universal function for systems whose underlying classical
dynamics is chaotic. Here $\langle \cdots \rangle$ denotes an average
over resonances $\nu$ and over different values of $\bar{X}$. Due to
the ergodic nature of the systems under investigation, one usually
conjectures that $\langle \cdots \rangle$ is equivalent to an average
over an ensemble of random Hamiltonians. The universality becomes
explicit after applying the rescalings
\begin{equation}
x = X \sqrt{C_v(0)} \; \; {\rm and} \; \; c_v(x) = C_v(X)/C_v(0)\;.
\label{eq:rescale}
\end{equation}
It is very difficult to measure $C_v(X)$ experimentally for quantum
dots and, consequently, to obtain the scale $C_v(0)$. The main goal of
this work is to address this difficulty by suggesting an alternative
way of studying parametric correlations. We show in Section IV that
one can estimate $C_v(0)$ by semiclassical arguments, once details of
the confining geometry of the dot are known. An important statement
implicit in Refs.~\onlinecite{Szafer93} and \onlinecite{Simons93a} is
that the quantity $\sqrt{C_v(0)}$ sets the scale for {\it any}
averaged parametric function $\langle f(X)\rangle$. Therefore, we also
study another (experimentally more accessible) function, the
conductance peak height correlator $C_g(X)$, which is defined as
\cite{Alhassid96,Bruus96}
\begin{equation}
C_g(X) \equiv \left\langle g_\nu ( \bar{X}-X/2 ) \, g_\nu (
\bar{X}+X/2 ) \right\rangle - \left\langle g_\nu ( \bar{X})
\right\rangle^2 \;.
\label{eq:defCX}
\end{equation}
This correlator also becomes universal upon rescaling $X$ according to
Eq.~(\ref{eq:rescale}) and $c_g(x) = C_g(X)/C_g(0)$. For the GOE case
an asymptotic expansion of $c_g(x)$ for $x\ll 1$ can, in principle, be
found analytically, in analogy to the GUE case shown in
Ref.~\cite{Bruus96}. Here, however, we only present results based on
numerical simulations, which nevertheless cover the {\it entire} $X$
range. For that purpose, we have performed a series of exact
diagonalizations of random matrices of the form $H(X) = H_1 \cos(X) +
H_2\sin(X)$, with $H_1$ and $H_2$ denoting two $500\times 500$
matrices drawn from the GOE. This model and the details of the
calculation are discussed in Ref.~\onlinecite{Bruus96}. The final
results for $c_v(x)$ and $c_g(x)$ are presented as full lines in
Figs.~\ref{fig:CvX} and \ref{fig:CgX}, respectively.

\section{Dynamical Model}
\label{sec:dynamical}

The aim of this section is to compare the statistical results of the
previous section with exact numerical diagonalizations of a dynamical
model. The essential characteristic of a dynamical model for this type
of study is a fair resemblance with the actual experimental
conditions, combined with its adequacy to numerical computations. For
this reason we chose the (two-dimensional) conformal billiard
\cite{Berry86,Bruus94,Bruus96}. Using complex coordinates, the shape
of the billiard in the $w$ plane is given by $|z|=1$ in the
area-preserving conformal mapping $w(z) = (z + bz^2 + c e^{i
\delta}z^3)/\sqrt{1+2b^2+3c^2}$, where $b$, $c$, and $\delta$ are real
parameters chosen in such manner that $|w'(z)| > 0$ for $|z| \leq
1$. In particular, we chose the values $b = c = 0.2$ and obtained
different boundary shapes by varying $\delta$. Hence, $X$ is now to be
identified with variations in $\delta$. The classical dynamics of a
particle bouncing inside billiards defined by these parameter values
is fully chaotic \cite{Bruus94}. The quantum dynamics is determined by
choosing Neumann (zero current) boundary conditions, which yield
non-trivial behavior at the boundary, as needed in
Eq.~(\ref{eq:gamma}). The eigenstates $\psi_{\nu}$ thus obtained
correspond to the resonant wave function appearing in
Eq.~(\ref{eq:gamma}) \cite{Bruus94}.
 
In practical calculations, one uses as a truncated basis the lowest
(in our case 1000) eigenstates of the circular billiard ($b=c=0$),
which are known analytically. To solve the Schr\"{o}dinger equation in
this basis one has to calculate several thousand matrix elements of
the Jacobian ${\cal J} = |w'(z)|^2$. However, using polar coordinate
separation, only matrix elements of powers of the radial coordinate
need to be calculated numerically. The angular part is done
analytically and changes in shape are trivial, since $b$, $c$, and
$\delta$ act as prefactors to the matrix elements. We have calculated
the spectrum for $0.41 \leq \delta \leq 2.73$ in steps of 0.04,
whereby the two values $\delta=0,\pi$ which lead to spatial symmetry
were avoided. Due to the truncation of the basis, only the lowest 250
of the calculated 1000 eigenstates were accurate enough to be used in
the analysis. We also discarded the lowest 150 eigenstates because of
their markedly nonuniversal behavior. The spectrum was then unfolded
using the Weyl formula. We emphasize that it is not a trivial task to
increase the number of converged states. No symmetry reduction of the
resulting eigenvalue problem is possible for the asymmetric conformal
billiard.

The scaling parameter $C_v(0)$ was calculated using Eq.~(\ref{CvXdef})
and averaging over groups of 20 levels and 12 shapes. For magnetic
flux variations we had found that $C_v(0)$ is proportional to
$\varepsilon^{1/2}$ \cite{Bruus96}. Here, for shape variations, we
find the following fit:
\begin{equation}
C_v(0)_{\varepsilon,\delta} = a_\delta\ \varepsilon^{3/2},
\label{eq:fit}
\end{equation}
with $\a_\delta = 0.0003 + 0.0014 \delta$. A physical explanation for
this energy dependence will be given in the next Section. We do not
have, however, a full undestanding of the $\delta$-dependence of
$a_\delta$ yet. Note that the numerical coefficients in $a_\delta$ are
specific to a billiard with an area equal to $\pi^2$.

We found that spectral fluctuations were rather large and for that
reason it seems more difficult to use shape than magnetic flux as the
external parameter to obtain good statistics, at least in numerical
simulations. Because of the $\delta$-dependence of $C_v(0)$, the
rescaling of $X$ has to be performed with care. The linear relation $x
= \sqrt{C_v(0)} X$ holds only locally. We therefore used the
generalized relation
\begin{equation}
x(\varepsilon,X) \equiv \int_{0}^{X}
\sqrt{C_v(0)_{\varepsilon,\delta}} \: d\delta.
\end{equation}
In Figs.~\ref{fig:CvX} and \ref{fig:CgX} we show the numerical results
for $c_v(x)$ and $c_g(x)$, respectively. They both compare rather well
with the curves obtained from the GOE matrix diagonalizations.

\section{Semiclassical estimate of $C_{\lowercase{v}}(0)$}
\label{sec:semicl}

In a recent paper \cite{Berry94}, Berry and Keating obtained an
approximate expression for the level velocity correlator
[Eq.~(\ref{CvXdef})] by combining derivatives of the cumulative level
density expressed semiclassically with the Gutzwiller trace
formula. Their calculation applied for the case where $X$ is an
Aharonov-Bohm flux line. In what follows we generalize the derivation
presented in Ref.\onlinecite{Berry94} for an arbitrary parametric
variation of the Hamiltonian. Furthermore, here it will become more
transparent why the result is universal for any chaotic system,
depending only on whether the time-reversal symmetry is preserved or
not. We do not consider the case of broken rotation symmetry
(symplectic ensemble).

The derivation presented in Ref.\onlinecite{Berry94} approximates the
velocity level correlator as the following function of an external
parameter $X$:
\begin{equation}
\label{Feta}
F_\eta (X, E) = \frac{1}{\delta\bar{X}}\int_{\delta\bar{X}} d\bar{X}
     \left\langle \frac{dN_\eta(E;\bar{X}-X/2)}{d\bar{X}}\
     \frac{dN_\eta(E;\bar{X}+X/2)}{d\bar{X}} \right\rangle_{\delta
     E}\;,
\end{equation}
where the average is taken over a window of width $\delta E\gg\Delta$
around $E$. One should view this approximation with a certain caution.
It is in fact exact for $X=0$ once the regularization parameter $\eta$
is gauged properly. For large values of $X$, Eq.~(\ref{Feta}) contains
the leading order term of $C_v(X)$. For any other (intermediate)
values of $X$, $F_\eta$ just {\it interpolates} smoothly between the
correct asymptotic values. The inaccuracy over intermediate values of
$X$ does not rely {\sl a priori} on the approximation used in the
evaluation of $F_\eta$, but has a trivial origin: One cannot express
$C_v(X)$ exactly by a 2-point, fixed energy correlator, as in
Eq.~(\ref{Feta}).

Equation~(\ref{Feta}) can be evaluated semiclassically by expressing
the cumulative level density $N_\eta$ by the trace formula $N_\eta(X;
E) = \sum_\mu A_\mu \exp (i S_\mu /\hbar - \eta T_\mu/\hbar)$, where
the sum runs over all periodic orbits. The amplitude $A_\mu$ contains
information about the stability of a certain orbit $\mu$, with $S_\mu$
being its action, and $T_\mu$ its period. All $A_\mu$, $S_\mu$, and
$T_\mu$ depend on $E$ and $X$. The coefficient $\eta$ acts as a
regulator of the infinite summation. Below, our approach will cover
only the range where the classical perturbation theory is valid,
which, nonetheless, can correspond to large changes in the spectrum,
depending on $\hbar$ \cite{Obs1}. With that in mind, we assume that a
shape distortion will only change the action of a periodic orbit by
$S_\mu (X) \approx S_\mu(X_0) + Q_\mu (X - X_0)$, with $Q_\mu =
dS_\mu(X_0)/dX$. The change in the orbit itself is assumed to be a
higher order effect, allowing us to neglect variations in $A_\mu$ as
compared to $Q_\mu/\hbar$. After some standard manipulations, the
diagonal term of $F_\eta$ can be written as
\begin{equation}
\label{Fdiag}
F^{diag}_\eta (X; E) = \frac{2}{\hbar^2} \left\langle \sum_\mu
   |A_\mu|^2 Q_\mu^2 \exp\left( -\frac{i}{\hbar} Q_\mu X -
   \frac{2\eta}{\hbar} T_\mu \right) \right\rangle_{\delta E} \;.
\end{equation}

Two further steps allow us to establish the connection between
Eq.~(\ref{Fdiag}) and the asymptotic part of level velocity
correlator. Due to the chaotic nature of the system, the longer the
orbits become, the more they explore ergodically the phase space. One
can assign a scale $T_{erg}$ to the time when periodic orbits with
period $T>T_{erg}$ begin covering uniformly the phase space. For
$T>T_{erg}$ the Hannay-Oz\'orio de Almeida sum rule \cite{Hannay84} is
valid, and the sum in Eq.~(\ref{Fdiag}) can be enormously simplified
\cite{Berry85}, since $\phi(T) = \left\langle \sum_\mu |A_\mu|^2
\delta(T-T_\mu)\right\rangle_{\delta T} \longrightarrow
1/2\pi^2|T|$. The average $\langle \dots \rangle$ is now taken over a
small window $\delta T$ which, nevertheless, contains a large number
of periodic orbits. Notice that this expression specializes our
results, since it is stated in a form suitable only for time-reversal
symmetric systems. (It is trivial to extend our results to systems
where this symmetry is absent.) In order to proceed analytically, one
has to extrapolate the Hannay-Oz\'orio de Almeida sum rule beyond its
range of applicability, i.e., for $T<T_{erg}$ \cite{Obs2}. This
approximation is not controllable, but its accuracy is supported by
numerical evidence \cite{Bruus96}. Hence,
\begin{equation}
\label{quasela}
F^{diag}_\eta(X; E) \approx \int_0^\infty \!dT \,\phi(T) \left\langle
Q_\mu^2 \exp\left( \frac{i}{\hbar} Q_\mu X\right)
\right\rangle_{\delta T} \;,
\end{equation}
with $\langle \cdots \rangle_{\delta T}$ now indicating an average
over periodic orbits within a window $\delta T$, but at a fixed energy
$E$. This average can be easily handled following the ansatz proposed
in Ref.\onlinecite{Goldberg91}, which reads
\begin{equation}
\label{Ansatz}
 P(Q) = \frac{1}{\sqrt{2\pi \langle Q^2 \rangle_{\delta T} }}
		   \exp\left(- \frac{Q^2} {2\langle Q^2
		   \rangle_{\delta T} }\right), \qquad \mbox{with}
		   \qquad \langle Q^2 \rangle_{\delta T} = \alpha(E) T
		   \;,
\end{equation}
where $P(Q)$ is the probability density of $Q$. Thus, $\alpha$ is a
system-dependent quantity which contains information about the
long-time dynamics, therefore probing some of its global (average)
properties. For billiards in the semiclassical regime, the energy
dependence of $\alpha$ follows straightforwardly from the fact that to
a trajectory of length $L$ corresponds an action $S=(2mE)^{1/2} L$,
yielding
\begin{eqnarray}
\langle Q^2(T)\rangle_{\delta T} & = & 2mE \int_0^T dt'
\int_0^T dt \left\langle \frac{d{\dot L}(t^\prime)}{dX} 
\frac{d{\dot L}(t)}{dX}
\right\rangle_{\delta T} \nonumber \\ & \approx & 2mET
\int_0^\infty dt\ C(t) \;.
\end{eqnarray}
This relation is similar to that obtained in Ref.\onlinecite{Bruus96}
for the variance of the winding number of classical
trajectories. Here, $d{\dot L}(t)$ gives the rate in time by which the
length of a trajectory changes under a variation $dX$ in the billiard
shape. For a chaotic system one expects not only that $\langle d{\dot
L}(t)/dX \rangle_{\delta T} = 0$ (because different trajectories are
uncorrelated), but also that the autocorrelator function of $d{\dot
L}(t)/dX$ decays sufficiently fast to make the integral over $C(t)$
finite. The magnitude of $\int dt C(t)$ is determined solely by the
velocity of the particle and the billiard area $\cal{A}$. By a proper
rescaling one can write $\alpha(E) = (8mE^3 {\cal A} )^{1/2} \kappa$,
where $\kappa$ is a classical, dimensionless quantity computed for an
unit area billiard for trajectories with unit velocity. The energy
dependence of $\alpha$ is markedly different from the case where the
magnetic field plays the role of the external parameter. Notice that
$\alpha(E)$ matches the largest contribution to $C_v(0)$ found in our
numerical fit [Eq~(\ref{eq:fit})].

A this point the semiclassical interpretation of universality becomes
apparent: The classical long-time dynamics of chaotic systems has some
universal properties which are the only necessary elements to evaluate
density-density correlators in the semiclassical approximation. In
addition to that, from the classical dynamics one can obtain
system-specific information which helps to understand relevant scales
for different experimental situations.

After replacing the average of the term involving $Q_\mu$ by an
integral over the distribution defined in Eq.~(\ref{Ansatz}), we
obtain
\begin{equation}
\label{F0}
F_{\eta=\Delta/2\pi}^{diag}(0; E) = \frac{\alpha(E)}{\pi\hbar\Delta}
		    \;,
\end{equation}
Equation~(\ref{F0}) gives a semiclassical estimate for $C_v(0)$. Upon
the usual rescaling [Eq.~(\ref{eq:rescale})], we get
\begin{equation}
f(x; E) = \frac{1 - \pi^2 x^2/2} {(1 +\pi^2 x^2/2)^2} \;,
\end{equation}
which, as expected, exhibits the correct large-$x$ asymptotic behavior
for time-reversal symmetric systems \cite{Simons93a}, namely, $f(x; E)
\rightarrow - 2/(\pi^2 x^2)$ for $x \gg 1$.

The classical evaluation of $\kappa$ makes it possible to predict
$C_v(0)$. For the conformal billiard, for instance, this implies in a
simulation of classical trajectories in slightly different
geometries. We emphasize that in order to give an accurate prediction
for $\kappa$ in experiments, it is necessary to know very precisely
the actual geometry of the quantum dot in consideration.

\section{Conclusions}
\label{sec:conclusions}

In this paper, we have argued that the universal form of the
parametric correlator of conductance peak heights indicates the
chaotic nature of the electron dynamics in quantum dots in the Coulomb
blockade regime. In the absence of magnetic fields, the simplest
parameter to vary experimentally is the shape of the quantum
dot. Whereas RMT provides the framework for obtaining the universal
form of the correlation function, the nonuniversal correlation scale
can be understood in simple semiclassical terms. This scale is rather
sensitive to the geometry of the dot and the Fermi energy.

We have compared a theoretical curve obtained from numerical
simulations of random matrices with the exact correlator obtained for
the conformal billiard after averaging over energy and shape
deformation. The agreement found was good, given the limitations
imposed by the size of our data set. We also found that the
semiclassical theory can explain the energy scaling of $C_v(0)$ on
very generic grounds -- shedding some light on how to make
quantitative predictions about nonuniversal characteristics of such
systems from a different (purely classical) perspective. We remark,
however, that shape variations are harder to treat classically, since
they alter trajectories strongly. To obtain fair estimates of the
scaling one needs to evaluate $\langle Q^2 \rangle_{\delta T}$ 
with great numerical accuracy and use very small increments in shape
deformation. Further investigation in this line is required to make a
firm statement. It is interesting to note, though, that the energy
dependence of the scaling is distinct for variations of boundary shape
or external magnetic field.



\newpage

\begin{thebibliography}{99}
\bibitem{Obs1} This is true because, for a billiard with
	$S_\mu=kL_\mu$, a small shape distortion (which originates a
	generic small variation $\delta L_\mu$ in the length of the
	$\mu$-th periodic orbit) can still change the spectrum
	dramatically, provided that $k\delta L_\mu/\hbar$ is at least
	of $O(1)$.
\bibitem{Obs2}
	We also assume that $Q_\mu$ and $A_\mu$ are not correlated,
	which is certainly true for $T>T_{erg}$.
\end{thebibliography}
\end{document}